\documentclass[submission, Proceedings]{SciPost}

\hypersetup{
    colorlinks,
    linkcolor={red!50!black},
    citecolor={blue!50!black},
    urlcolor={blue!80!black}
}

\usepackage{CJKutf8}
\usepackage{amsfonts}
\usepackage{setspace,lipsum}

\usepackage[bitstream-charter]{mathdesign}
\DeclareSymbolFont{usualmathcal}{OMS}{cmsy}{m}{n}
\DeclareSymbolFontAlphabet{\mathcal}{usualmathcal}

\begin{document}

\begin{center}{\Large \textbf{
Light-cone PDFs and GPDs from Lattice QCD\\
}}\end{center}

% TODO: write the author list here. Use initials + surname format.
% Separate subsequent authors by a comma, omit comma at the end of the list.
% Mark the corresponding author with a superscript *.
\begin{center}
Martha Constantinou\textsuperscript{1$\star$}
\end{center}

\begin{center}
{\bf 1} Temple University, Philadelphia, USA
\\
* marthac@temple.edu
\end{center}

\begin{center}
\today
\end{center}

% For convenience during refereeing (optional),
% you can turn on line numbers by uncommenting the next line:
%\linenumbers

\definecolor{palegray}{gray}{0.95}
\begin{center}
\colorbox{palegray}{
  \begin{tabular}{rr}
  \begin{minipage}{0.1\textwidth}
    \includegraphics[width=30mm]{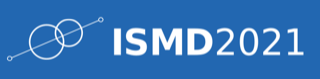}
  \end{minipage}
  &
  \begin{minipage}{0.75\textwidth}
    \begin{center}
    {\it 50th International Symposium on Multiparticle Dynamics}\\ {\it (ISMD2021)}\\
    {\it 12-16 July 2021} \\
    \doi{10.21468/SciPostPhysProc.?}\\
    \end{center}
  \end{minipage}
\end{tabular}
}
\end{center}

\section*{Abstract}
{\bf
% TODO: write your abstract here.
In this article, we review recent lattice calculations on the $x$-dependence of PDFs and GPDs from lattice QCD.
}

\vspace{5pt}
\noindent\rule{\textwidth}{1pt}
\tableofcontents\thispagestyle{fancy}
\noindent\rule{\textwidth}{1pt}
\vspace{5pt}

The QCD factorization formalism is the foundation for understanding hadrons in terms of their partonic constituents. The partonic structure is mapped via the parton distribution functions (PDFs), the generalized parton distributions (GPDs), and the transverse-momentum-dependent distributions (TMD PDFs). The PDFs, GPDs, and TMD PDFs are light-cone correlation functions and cannot be accessed from the Euclidean formulation of lattice QCD, but partial information can be obtained from their Mellin moments.
Alternative approaches to Mellin moments have been pursued and here we focus on the quasi-PDFs approach via Large-Momentum Effective Theory (LaMET)~\cite{Ji:2013dva}, and the pseudo-PDFs~\cite{Radyushkin:2016hsy}, which is based on short-distance factorization.
The progress in lattice QCD calculations on the $x$-dependence of PDFs GPDs and even TMDs has been impressive. Here, we use some of the pioneering contributions as presented in the review talk. For extensive reviews, we direct the reader to Refs.~\cite{Cichy:2018mum,Constantinou:2020pek}. 

%###########################################
%###########################################
\vspace*{-0.5cm}
\section{PDFs}
\label{sec:PDFs}

PDFs are easier to obtain compared to GPDs and TMDs, as they are one-dimensional objects depending only on the momentum fraction carried by the partons. Their twist classification indicates the order at which a distribution appears in the expansion in terms of the large energy, $Q$, of the physical process.  At leading twist, PDFs are the unpolarized, helicity, and transversity, and appear in different physical processes making their extraction subject to different challenges. To date, there is still missing information, either because it is challenging to extract certain PDFs or because some kinematic regions are not easily accessible. Therefore, lattice QCD input is crucial, and lattice results can either serve as predictions or as additional constraints on the phenomenological analyses. Here we discuss the current landscape of lattice results, highlighting their $x$-dependence in the presentation, rather than technical details.

\bigskip
The majority of lattice calculations is on PDFs and the progress led to calculations at the physical point. Currently, the focus is on understanding and quantifying systematic uncertainties. In these proceedings we focus on nucleon PDFs using simulations at the physical point, and we highlight calculations at the physical point obtained from the quasi and pseudo PDFs methods. Both approaches demonstrated that matrix elements of boosted hadron states coupled with bilinear non-local operators containing a Wilson line in the direction of the boost,
\begin{equation}
{\cal M}(P_3,z,\mu) = Z_\Gamma(z,\mu) \langle H(P_3) | \bar{\psi}(z)\Gamma W(z) \psi(0) |H(P_3) \rangle\,,
\label{eq:ME}
\end{equation}
 can be related to light-cone distributions. Naturally, these matrix elements are defined in coordinate space, with the length of the Wilson line being varied from zero up to half the spatial extent of the lattice. The Dirac structure is $\Gamma=\gamma^0,\,\gamma^3\gamma^5,\,\sigma^{3j}$ ($j=1,2$) for the unpolarized, helicity and transversity PDFs, respectively. $Z_\Gamma(z,\mu)$ is the renormalization function defined at renormalization scale $\mu$. The $x$ dependence of the quasi-PDFs is reconstructed via a Fourier transform, FT, (or advanced reconstruction techniques, such as Backus-Gilbert, Bayesian methods) of ${\cal M}(P_3,z,\mu)$. For a large enough momentum, the quasi-PDFs are related to the light-cone PDFs via a matching procedure in large momentum effective theory (LaMET). The pseudo-PDFs method relies on the reduced Ioffe time pseudo-distribution, that is
\begin{equation}
\label{eq:reduced}
\mathfrak{M}(\nu,z^2) =
\frac{\mathcal{M}(\nu,z^2)\,/\,\mathcal{M}(\nu,0)}{\mathcal{M}(0,z^2)\,/\,\mathcal{M}(0,0)}
\end{equation}
$\nu\equiv z\cdot P_3$ is the so-called Ioffe time, and $\mathfrak{M}(\nu,z^2)$ is calculated for a wide range of $\nu$ by varying $P_3$, or $z$, or both. $\mathfrak{M}(\nu,z^2)$ is matched to the light-cone ITDs using a matching procedure based on short-distance expansion. The PDFs are related to ITDs via an FT in Ioffe time.

\begin{figure}[h!]
\begin{minipage}{9cm}
\hspace*{-0.15cm}
\includegraphics[scale=0.75]{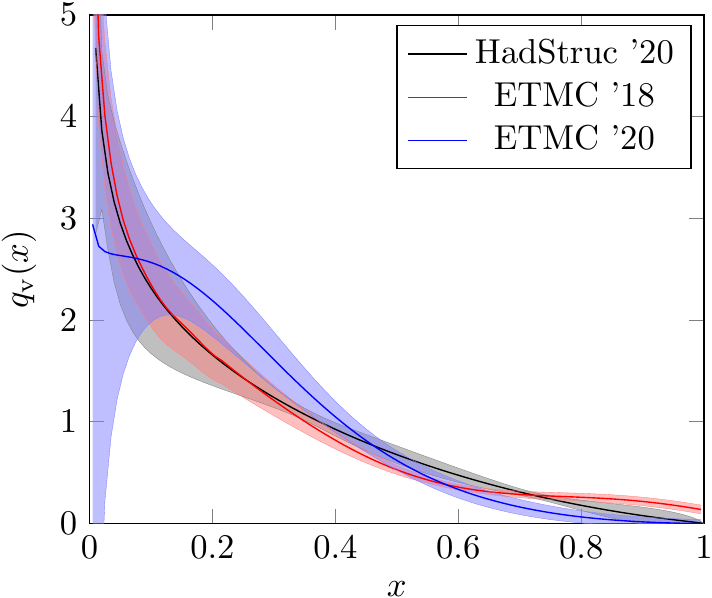}\hspace*{0.1cm}
\includegraphics[scale=0.21]{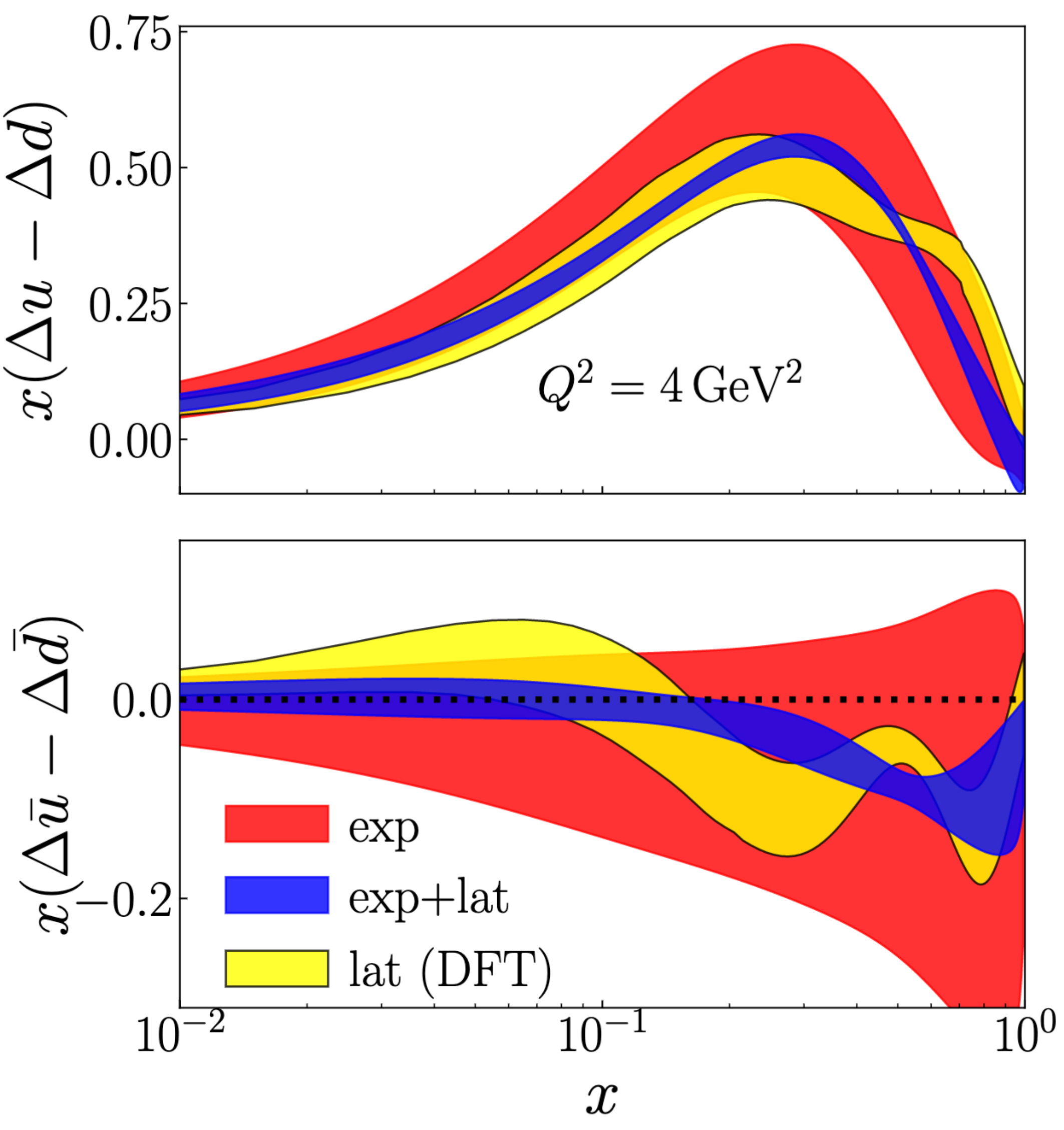}
\end{minipage}
\hfill
\hspace*{5.5cm}
\begin{minipage}{5cm}
\caption{\footnotesize{Right: Lattice QCD results\\ on the unpolarized PDF using the \\  quasi-PDFs method~\cite{Alexandrou:2019lfo}  (red band) \\ and
pseudo-ITDs from Ref.~\cite{Joo:2020spy} (gray \\ band) and Ref.~\cite{Bhat:2020ktg} 
(blue band). \\ Right: The isovector helicity PDF. \\
The JAM17 results are shown  \\ with  red 
band, while the blue band \\ shows the combined fits using both \\ lattice and  experimental data. The \\ yellow band  shows the data from \\  Refs.~\cite{Alexandrou:2018pbm}. Source: Refs.~\cite{Joo:2020spy,Bringewatt:2020ixn}.}}
\label{fig:nucleon_QPFG_pPDF}
\end{minipage}
\vspace*{-0.4cm}
\end{figure}

Results on the unpolarized PDF using ensembles at physical quark masses can be found in Refs.~\cite{Alexandrou:2018pbm,Joo:2020spy,Bhat:2020ktg}, and are shown in Fig.~\ref{fig:nucleon_QPFG_pPDF}. The study of Ref.~\cite{Joo:2020spy} corresponds to different lattice discretization than Refs.\cite{Alexandrou:2018pbm,Bhat:2020ktg}. Also, Ref.~\cite{Alexandrou:2018pbm} uses the quasi-PDFs method, while the other two implement the pseudo-PDFs. The results are very encouraging, exhibiting agreement for a wide range of values for $x$. However, there is some tension at large $x$, which may potentially be explained by systematic effects. The success in extracting the $x$ dependence of PDFs is a significant achievement for lattice QCD, and has the potential to help constrain PDFs in kinematic regions where experimental data are not available. Studies exploring the synergy of lattice QCD results and global analysis can be found in Refs.~\cite{Cichy:2019ebf,Bringewatt:2020ixn,DelDebbio:2020rgv}. In the right panel of Fig.~\ref{fig:nucleon_QPFG_pPDF} we show a combined analysis of the helicity PDF using the lattice data from Ref.~\cite{Alexandrou:2018pbm} and experimental data sets within the JAM framework, performed in Ref.~\cite{Bringewatt:2020ixn}. The effect of the lattice data in the unpolarized PDFs does not further constrain the PDFs. In contrast, the lattice data on the helicity PDFs provide further constraints to the combined estimates, as can be seen in Fig.~\ref{fig:nucleon_QPFG_pPDF}. It is found that the combined estimates of the quark PDFs have reduced uncertainties by a factor of 3 in the small- and large-$x$ regions and a factor of 5 in the intermediate-$x$ regions. The anti-quark isovector helicity PDF exhibits an equally large improvement.

Having results at physical values of the quark masses, the attention of the calculation has turned to investigation of systematic uncertainties. Ref.~\cite{Alexandrou:2019lfo}  studies selected sources of systematic effects, such as excited-states effects, concluding that lattice data at $T_{\rm sink}$ below 1 fm are insufficient to suppress excited-states, particularly as $P_3$ increases. Other sources of uncertainties studied are related to the renormalization, and the reconstruction of the $x$-dependence via a FT and the matching procedure. Finite-volume effects have been studied for the unpolarized and helicity PDFs in Ref.~\cite{Lin:2019ocg}. The matrix elements from different ensembles were found to be compatible at the same boost, indicating that volume-effects are not significant for $m_{\rm val} L \in [3.3 - 5.5]$. No significant volume effects were found in Ref.~\cite{Fan:2020nzz}. However, excited states contamination increase as $P_3$ increases. More recently, studies of discretization effects became available, which is important because nonlocal operators are not ${\cal O}(a)$ improved. Ref.~\cite{Alexandrou:2020qtt} studies the unpolarized and helicity PDFs at a pion mass of about 370 MeV and three values of the lattice spacing (0.064 - 0.093 fm). Ref.~ \cite{Karpie:2021pap} uses the pseudo-PDFs approach to extract the unpolarized PDF with an ensemble of 400 MeV and three lattice spacings (0.048 - 0.075 fm). 
Ref.~\cite{Lin:2020fsj} uses ensembes at three lattice spacings (0.058 - 0.12 fm) and three values of the pion masses. 
In a nutshell, it is found from the various calculations that discretization effects are rather large, and a continuum extrapolation is currently challenging. 

Another highlight of the last year is the flavor decomposition of the unpolarized, helicity, and transversity PDFs, which includes the calculation of disconnected contributions. To date, there are only two calculations~\cite{Zhang:2020dkn,Alexandrou:2020uyt}, where the mixing with the gluon PDFs is neglected. Ref.~\cite{Zhang:2020dkn} presents only the strange and charm unpolarized PDFs , which are less costly than the light quark contributions, and use two values of the valence pion mass, 310 and 690 MeV. The first up- and down-quark PDFs have been obtained in Ref.~\cite{Alexandrou:2020uyt} for the helicity, followed by the longer write-up of Ref.~\cite{Alexandrou:2021oih} for all three types of PDFs. The calculation is performed using one ensemble at pion mass 260 MeV.

\medskip
While twist-3 PDFs lack a probabilistic interpretation, they are very interesting, as they contain information on the soft dynamics, such as the transverse polarization. Also, they are not smaller in magnitude than the twist-2 PDFs, and thus, are important to be considered. It is very challenging to probe them experimentally and isolate them from the leading-twist contributions. There are two chiral odd twist-3 PDFs, $e(x)$ and $h_L(x)$, and one chiral even that is $g_T(x)$. Lattice results on $g_T$ and $h_L$ have appeared recently~\cite{Bhattacharya:2020cen,Bhattacharya:2021moj}.

\begin{figure}[h!]
\vspace*{-0.35cm}
\begin{minipage}{11cm}
\hspace*{-0.15cm}
\includegraphics[scale=0.39]{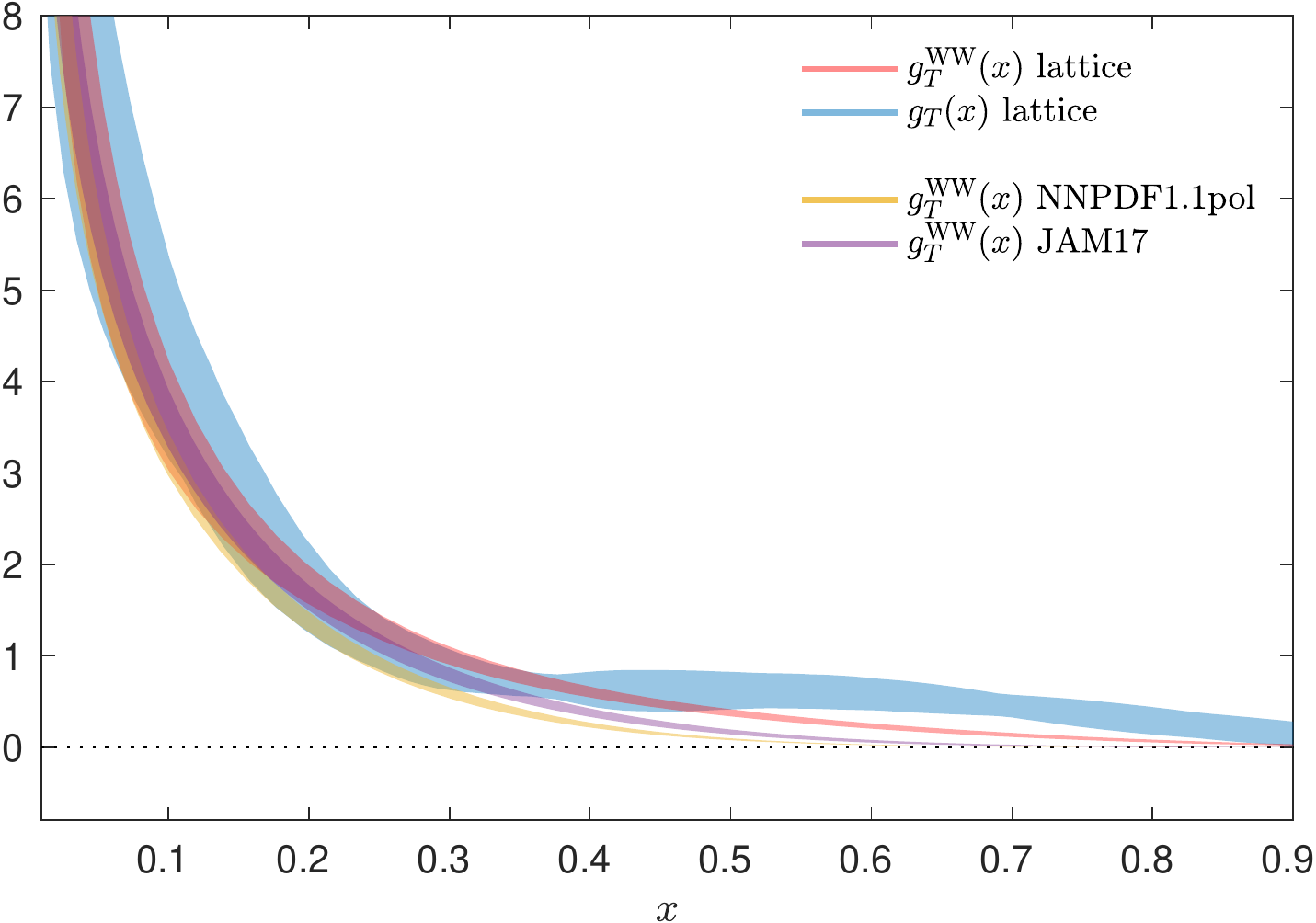}
    \includegraphics[scale=0.39]{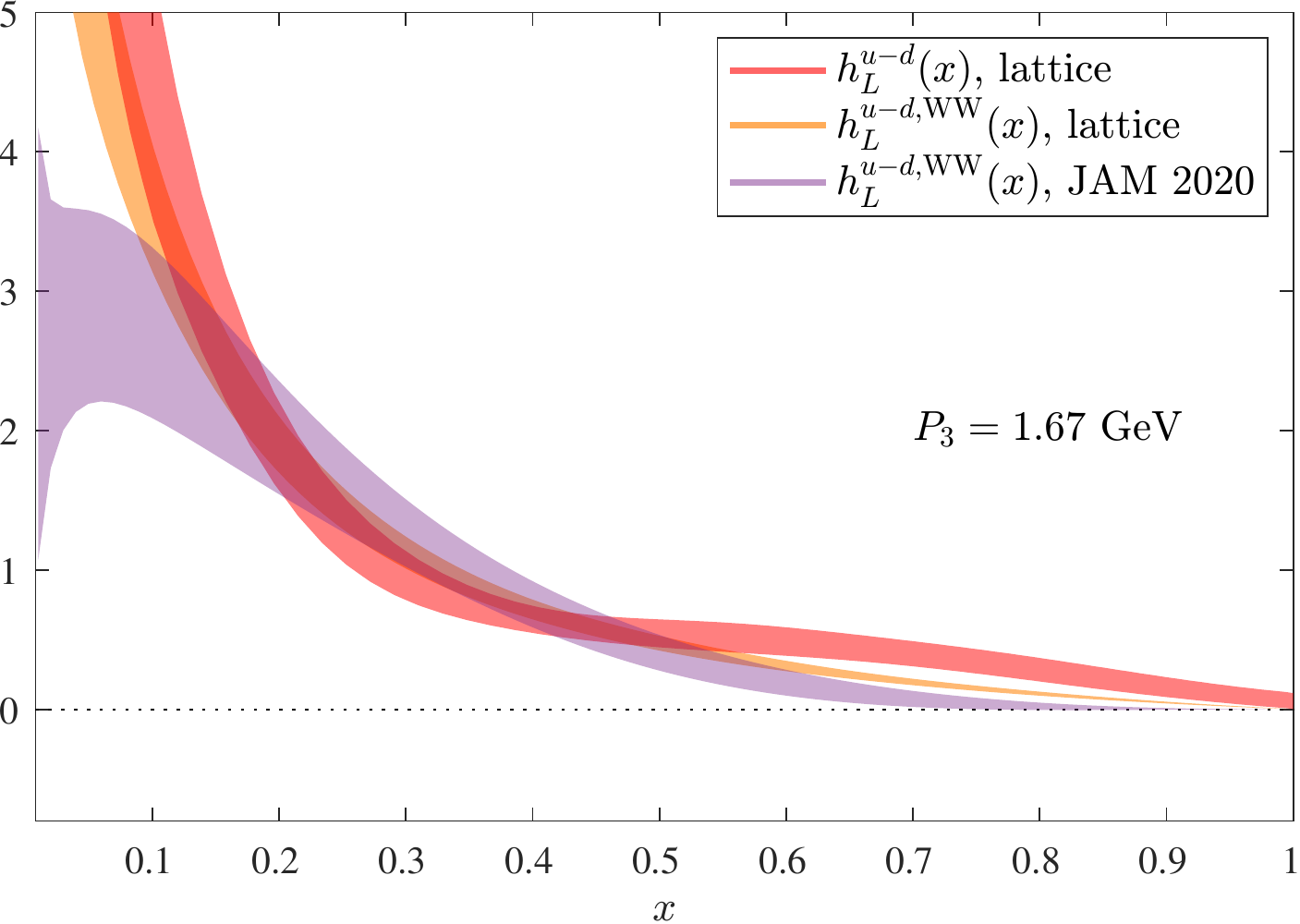}
\end{minipage}
\hfill
%\hspace*{5.5cm}
\begin{minipage}{3cm}
	\caption{\footnotesize{Comparison  \\ of  $g_T(x)$ (left) and $h_L(x)$ \\  (right) at  $P_3=1.67$ GeV \\  with their  WW  lattice  \\ estimates. WW estimates \\ using global fits data  \\ are shown and references \\ are in the original \\ publications.}}
	 	\label{fig:WW_approx}
\end{minipage}	
\end{figure}

The twist-3 PDFs are connected to their twist-2 counterparts through the Wandzura-Wilczek (WW) approximation. In this approximation, $g_T(x)$ and $h_L(x)$ are fully determined by the twist-2 $g_1(x)$ and  $h_1(x)$, respectively, and quark-gluon correlations are neglected. The WW approximation for both $g_T(x)$ and $h_L(x)$ is shown in Fig.~\ref{fig:WW_approx}. It is found that the actual lattice data for $g_T(x)$ are consistent with $g_T^{\rm WW}(x)$ for a considerable $x$-range. However, given the uncertainties, a violation of the WW approximation is still possible at the level of up to 40\% for $x{\lesssim}0.4$. The results of $h_L(x)$ and $h^{\rm WW}_L(x)$ are in agreement for $x{\lesssim}0.55$. The lattice results on $h^{\rm WW}_L(x)$ in the region $0.15 \lesssim x{\lesssim}0.55$ are also in good agreement with $h^{\rm WW}_L(x)$ obtained from a global fit of the nucleon transversity by the JAM collaboration. These findings seem to suggest that $h_L(x)$ could be determined by the $h_1(x)$ for a considerable $x$-range. 

%\newpage
%

%###########################################
%###########################################
\vspace*{-1.5cm}
\section{GPDs}
\label{sec:GPDs}

\begin{figure}[h!]
\centering
    \includegraphics[scale=0.39]{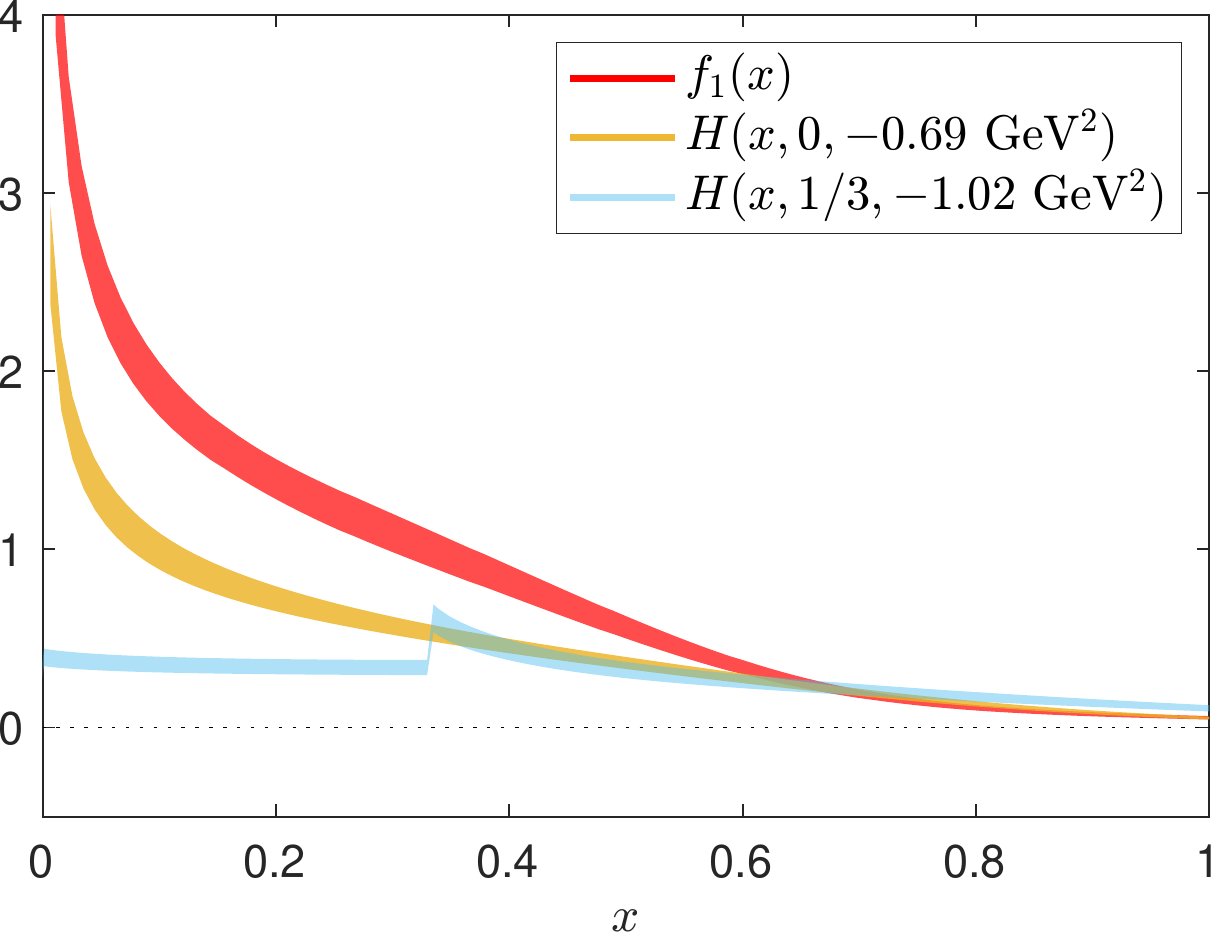} \includegraphics[scale=0.39]{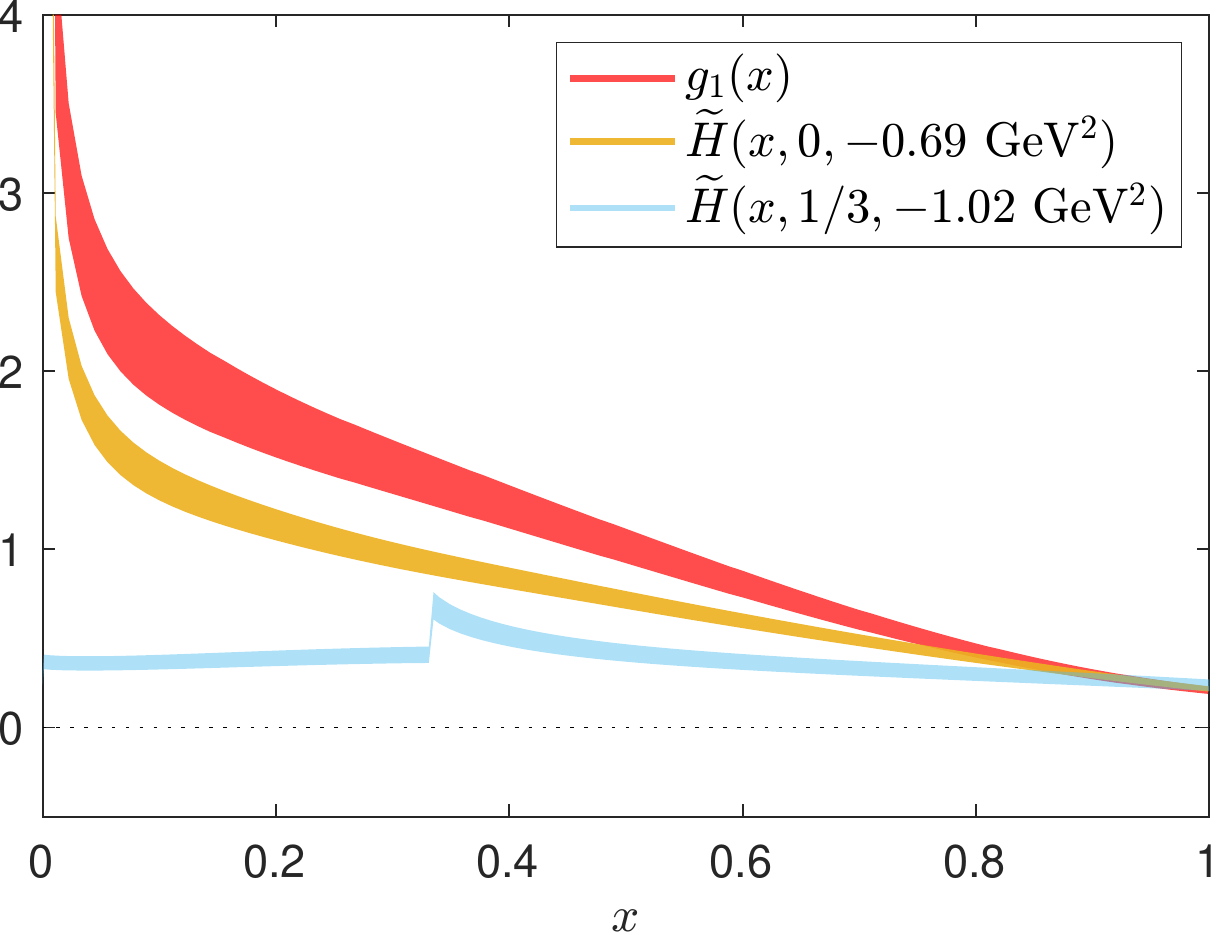} \includegraphics[scale=0.39]{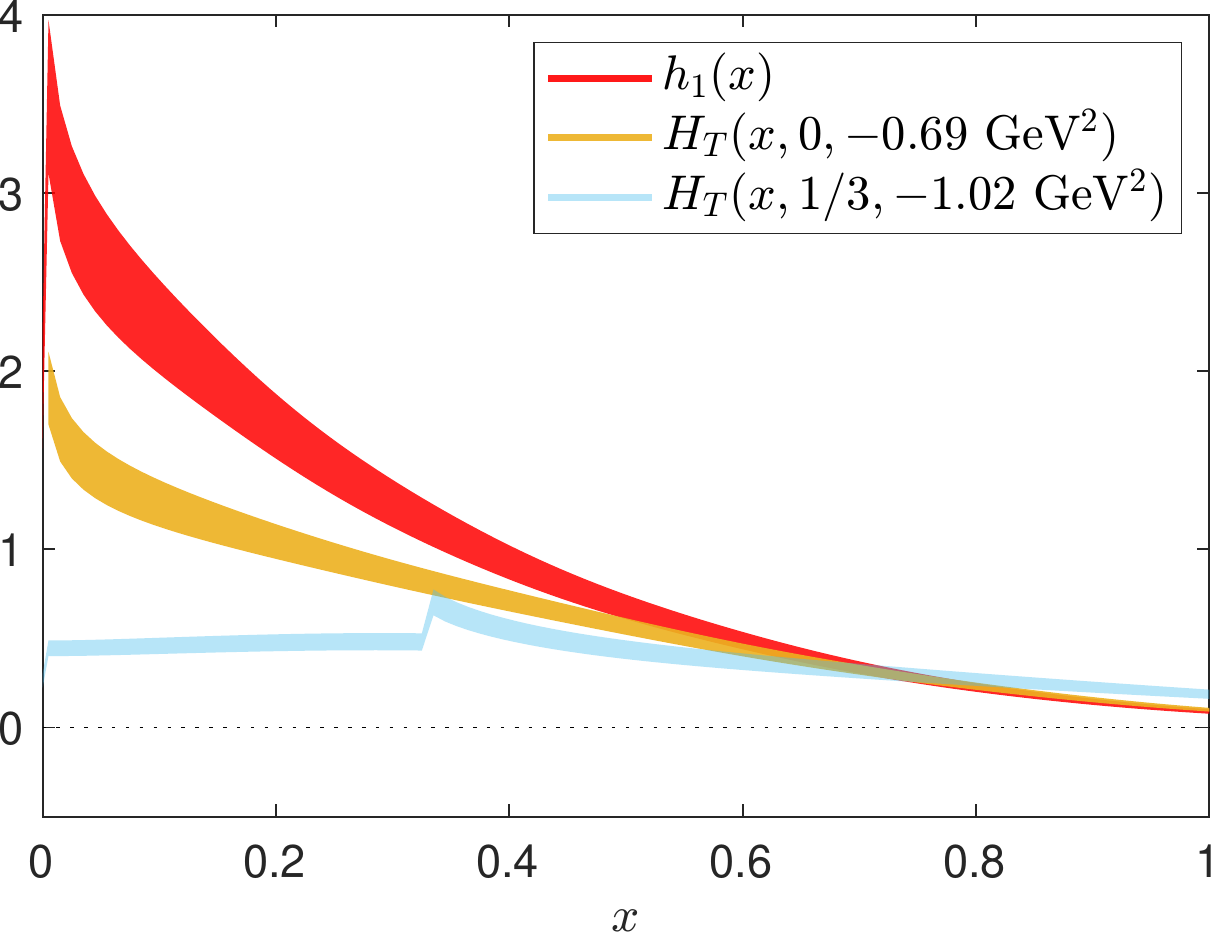}
    \vskip -0.25cm
    \caption{\footnotesize{The unpolarized $H$, helicity $\widetilde{H}$ and transversity $H_T$ GPDs at $\{t,|\xi|\}=\{0,0\},\,\{-0.69{\,\rm GeV}^2,0\},\,\{-1.02{\,\rm GeV}^2,1/3\}$ extracted from Ref.~\cite{Alexandrou:2020zbe,Alexandrou:2021bbo}.}}
    \label{fig:latt_GPDs}
\end{figure}

\vspace*{-0.25cm}
GPDs are relevant to non-forward scattering processes and depend on two additional kinematic variables in addition to $x$, that is, the light-cone component of the longitudinal momentum transfer (skewness), $\xi$, and the momentum transfer squared $t=\Delta^2$. Consequently, the GPDs are multi-dimensional quantities and their extraction is more challenging due to limited experimental data sets that cover a small kinematic region. The $x$-dependence of GPDs for fixed values of $t$ and $\xi$ can be accessed using the aforementioned methods. This new direction has recently been explored for the unpolarized, helicity and transversity proton GPDs~\cite{Alexandrou:2020zbe,Alexandrou:2021bbo} for an ensemble at pion mass 260 MeV, and a summary is presented in Fig.~\ref{fig:latt_GPDs}. Results on $H$, $\widetilde{H}$ and $H_T$ are shown for both zero and nonzero skewness ($\xi=\pm 1/3$) and are compared with their forward limit.
There is a qualitative difference between the GPDs at zero and nonzero skewness, due to the nontrivial ERBL region for $\xi\ne0$. It should be emphasized that such calculations are exploratory and are available for only a few values of $t$. Presently, lattice results are useful for a qualitative understanding of GPDs. For instance, one can find characteristics for the $t$ dependence for each operator under study.

\bigskip
In this review we highlight some of the many successes of the field of $x$-dependent distributions. Regardless of the necessary improvements, there is no doubt that lattice QCD calculations are progressing toward understanding the complex structure of hadrons.

\paragraph{Funding information}
The author acknowledges financial support by the U.S. Department of Energy, Office of Nuclear Physics, Early Career Award under Grant No.\ DE-SC0020405.

\begin{spacing}{0.6}
\bibliography{references.bib}
\end{spacing}

\nolinenumbers

\end{document}